\documentclass[twocolumn,superscriptaddress,english,prl,showpacs,longbibliography]{revtex4-2}
\usepackage[colorlinks=true,urlcolor=blue,citecolor=blue,linkcolor=blue]{hyperref} 

\usepackage{amsmath}
\usepackage{graphicx}% Include figure files
\usepackage{bm}% bold math
\usepackage{hyperref}% add hypertext capabilities
\usepackage{color}
\usepackage{amssymb}
\usepackage{graphicx}
\usepackage{color}
\usepackage{mathrsfs}
\usepackage{float}
\usepackage{indentfirst}
\usepackage{txfonts}
\usepackage{algorithm}  
\usepackage{algpseudocode}  
\usepackage{balance}
\usepackage[keeplastbox]{flushend}
\newcommand{\s}{\mathbf {s}}
\newcommand{\xeb}{F_{\mathrm{XEB}}}
\newcommand{\vv}{\mathbf {v}}
\newcommand{\head}{\mathrm{head}}
\newcommand{\ghead}{\mathcal G_{\mathrm{head}}}
\newcommand{\tail}{\mathrm{tail}}
\newcommand{\gtail}{\mathcal G_{\mathrm{tail}}}

\begin{document}

%\preprint{APS/123-QED}

\title{Simulating the Sycamore quantum supremacy circuits }% Force line breaks with \\
% \thanks{A footnote to the article title}%

\author{Feng Pan}
\affiliation{
% CAS Key Laboratory for Theoretical Physics, Institute of Theoretical Physics, Chinese Academy of Sciences, Beijing 100190, China
Institute of Theoretical Physics, Chinese Academy of Sciences, Beijing 100190, China
}
\affiliation{
 School of Physical Sciences, University of Chinese Academy of Sciences, Beijing 100049, China
}
\author{Pan Zhang}
\email{panzhang@itp.ac.cn}
\affiliation{
 %CAS Key Laboratory for Theoretical Physics, Institute of Theoretical Physics, Chinese Academy of Sciences, Beijing 100190, China
 Institute of Theoretical Physics, Chinese Academy of Sciences, Beijing 100190, China
}

%\date{\today}% It is always \today, today,
             %  but any date may be explicitly specified

\begin{abstract}
We propose a general tensor network method for simulating quantum circuits. The method is massively more efficient in computing a large number of correlated bitstring amplitudes and probabilities than existing methods.
As an application, we study the sampling problem of Google's Sycamore circuits, which are believed to be beyond the reach of classical supercomputers and have been used to demonstrate quantum supremacy. Using our method, employing a small computational cluster containing $60$ graphical processing units (GPUs), we have generated one million correlated bitstrings with some entries fixed, from the Sycamore circuit with $53$ qubits and $20$ cycles, with \textit{linear cross-entropy benchmark} (XEB) fidelity equals $0.739$, which is much higher than  those in Google’s quantum supremacy experiments. 
\end{abstract}
\maketitle
%\section{Introduction}
An essential question in near-term quantum computation is whether programmable quantum devices are able to beyond the ability of classical computations in specific computational tasks. An ideal example is sampling from a random quantum circuit~\cite{arute2019quantum,aaronson2016complexity,bouland2019complexity,boixo2017simulation,aaronson2019classical,zlokapa2020boundaries,boixo2018characterizing}. In 2019, Google's quantum computing group released the Sycamore circuits~\cite{arute2019quantum} with $n=53$ qubits, and demonstrated the \textit{quantum supremacy}~\cite{preskill2012quantum,boixo2018characterizing} by showing that they can experimentally solve the noisy sampling task from the output distribution $P_U(\s)$ of the Sycamore circuit $U$ in the computational basis $|\s\rangle$, which would cost $10,000$ years on modern supercomputers.

However, despite the great success of the experiments, we note that the critical basis for the assertion of the quantum supremacy, the accuracy of sampling in terms of the linear cross-entropy benchmark (XEB), and the running time of classic simulations, still leave some space for further discussions.
First, Google was only able to compute the exact XEB values for circuits up to $m=12$ cycles and estimated the XEB values from $14$ cycles to $20$ cycles using extrapolations~\cite{arute2019quantum}. This means that the fidelity of samples generated from the quantum supremacy circuits with $m\ge 14$ cycles are not verified. Second, the estimate of computational time was based on a classic simulation method, Schr\"odinger-Feynman algorithm \cite{aaronson2016complexity,markov2018quantum,arute2019quantum}. Potentially there could be other algorithms which are more efficient then the algorithm used by Google, demanding much less computational time than the estimate.

There are basically two kinds of methods for simulating quantum circuits. The first kind stores and evolves the full quantum state vector $\psi$, and is known as the Schr\"odinger method. Since bitstring probabilities $P_U(\s)=|\langle \psi| s\rangle|^2$ are known, sampling from the bitstring space is easy. The advantage of the method is that the computational complexity of the algorithm is linear in the number of depth $m$, thus it is very efficient for quantum circuits with a small number of qubits. Google used this method for simulating circuits up to $43$ qubits~\cite{arute2019quantum}. However, for a large number of qubits, this method suffers from exponential space complexity. Using a supercomputer, the largest instance that has been simulated has $49$ qubits~\cite{li2019quantum}, beyond which the size of total random access memory (RAM) becomes the bottleneck even with supercomputers. IBM has justified theoretically that the $53$-qubit state vector of the Sycamore circuits can be stored and evolved if one could employ not only all the RAM but also all the hard disks of the Summit supercomputer. However, the experiment has not been done yet.
%,luo2020yao fatima2020faster,
%\subsection{Existing methods and their limitations in simulating the Sycamore circuits} 
To address the issue of exponential space-complexity with a large number of qubits, Google used the Schr\"odinger-Feynman method~\cite{aaronson2016complexity,markov2018quantum} which breaks the circuits into two parts, connected using Feynman path-integrals, and each part is simulated using the Schr\"odinger method. Based on this method, Google estimated that simulating the Sycamore circuits with $20$ cycles requires 10,000-year running time on the Summit supercomputer.  

The second kind of method does not store all the $2^n$ bitstring probabilities in memory, but computes one bitstring probability or a small batch of them based on tensor networks ~\cite{markov2008simulating,gray2020hyper,guo2019general,chen2018classical,markov2018quantum,boixo2017simulation,boixo2018characterizing,pan2020contracting,villalonga2019flexible,villalonga2020establishing,huang2020classical,schutski2020simple}. Quantum circuits can be treated as particular tensor networks with unitary constraints. Given an initial state and the bitstring representing the measurements at the end, contraction of the tensor network gives the amplitude of the output bitstring. The space complexity of the tensor network method is controlled by the size of the largest tensor encountered during the contraction, which equals the exponential of tree width of the line graph corresponding to the tensor network~\cite{markov2008simulating}. For shallow circuits where the tree width is small, tensor network method is very efficient even for circuits with a large number of qubits~\cite{boixo2017simulation,chen2018classical,guo2019general,pan2020contracting}. However, the tensor network method is not scalable with the circuit depth, because the complexity is usually exponential to the depth hence very expensive for large circuits with sufficient depth. For the Sycamore circuits with $20$ cycles, a recent work~\cite{huang2020classical} (which is considered as the state-of-the-art) estimated that computing probabilities for a batch of $64$ bitstrings requires about $833$ seconds using a Summit-compatible supercomputer. To the best of our knowledge, so far there is no work that has successfully obtained the probability of even one bitstring for the Sycamore circuit with $20$ cycles, in practice.
More seriously, sampling from the bitstring space is difficult for the tensor network method, because the computational complexity is proportional to the number of samples one demands. In order to obtain enough bitstrings, e.g. for reaching an XEB value that is comparable to Google's hardware samples, the tensor network contraction has to be repeated for many times, making the overall computations intractable. 
%The problem of existing tensor network methods is that single or small-batch contraction already requires a large amount of computational resources, and the resources requires is proportional to the number of samples one need. [Take Ali's paper to illustrate]. As a conclusion, current methods is not able to produce a large number of samples.

In this article, we propose a tensor network method that enables us to fill the gap. Our method is based on the careful design of subspace to enumerate (or sample) from. The method can be regarded as an intermediate between the full-state-vector method and the single(or small-batch)-amplitude method. It is much more efficient than the Schr\"odinger-Feynman algorithm and the existing tensor-network methods for computing probabilities of a large number of correlated samples from the Sycamore circuits. 
%The computational cost of computing $2$ million bitstring probabilities is lower than that of computing $64$ bitstring probabilities of the state-of-the-art tensor network method~\cite{huang2020classical}. 
The computational complexity for obtaining $2$ million bitstring probabilities is much lower than the method used by Google~\cite{arute2019quantum}, and is also lower than the computational complexity for obtaining only $64$ amplitudes in the state-of-the-art tensor network method~\cite{huang2020classical}. 
Our method enjoys massive parallelization over multiple GPUs. By employing a small computational cluster composed of $60$ GPUs, we have computed exact amplitudes and probabilities for $2$ million correlated bitstrings in $5$ days from the Sycamore circuits with $53$ qubits and $20$ cycles. We have obtained $1$ million bistrings from a bistring sub-space, achieving XEB fidelity $0.739$, which is much higher than the samples obtained in Google's experiments.

\section{Big-head simulation of quantum circuits}
Storing the full state-vector has the advantage of easy sampling for a large number of bitstrings because the probability distribution, $P_U(\s)$, is stored in memory. But it suffers from the large space complexity. Computing once a time the probability of a single amplitude (or a small batch of amplitudes) using tensor network contractions largely reduces the computational complexity. However, the sampling task is difficult because one has to repeat the tensor network contraction process. Our idea is to combine the advantages of two methods, using a sub-space simulation. 

\begin{figure}[h]
\centering
\includegraphics[width=0.9\columnwidth]{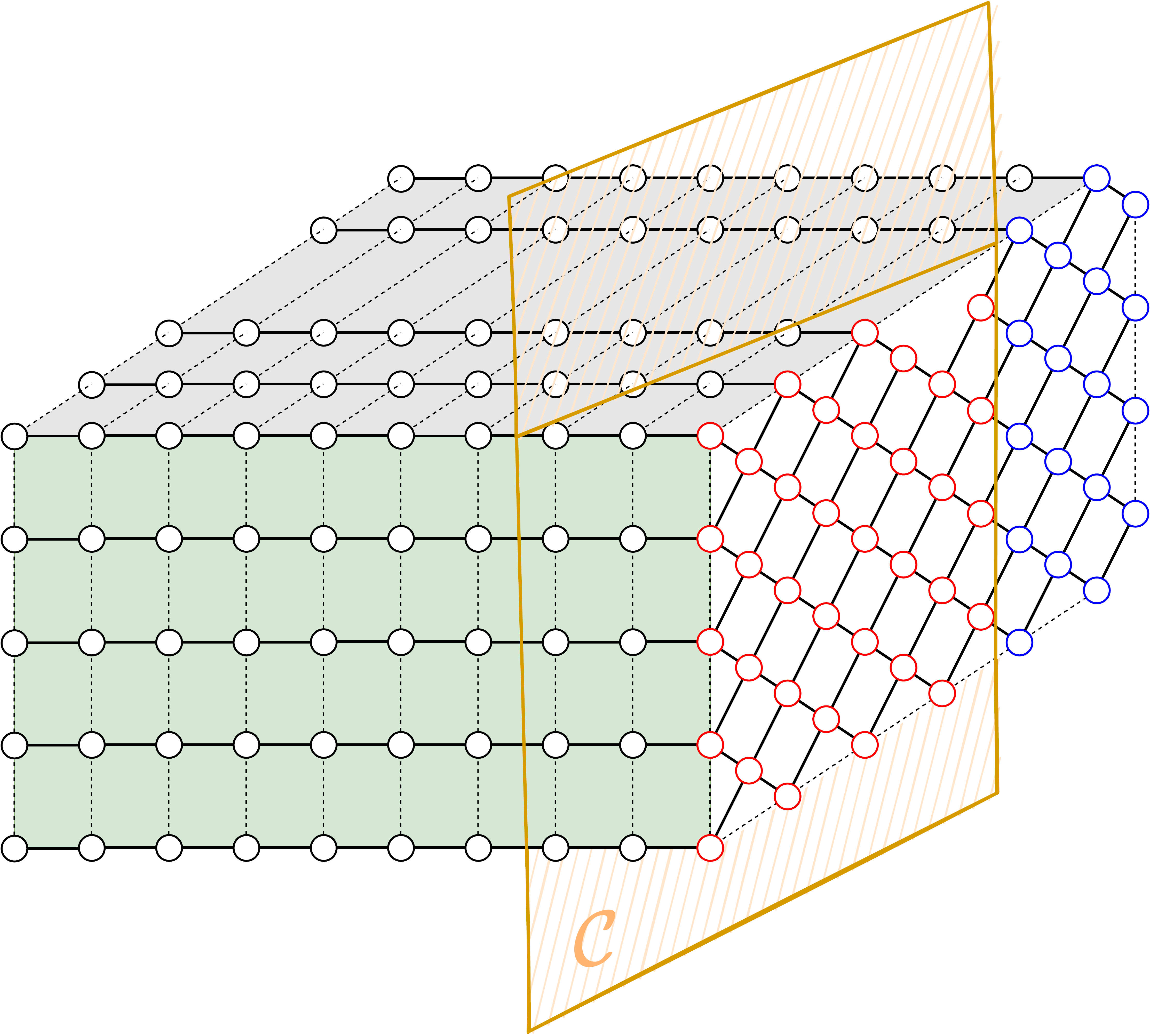}
\includegraphics[width=1.0\columnwidth]{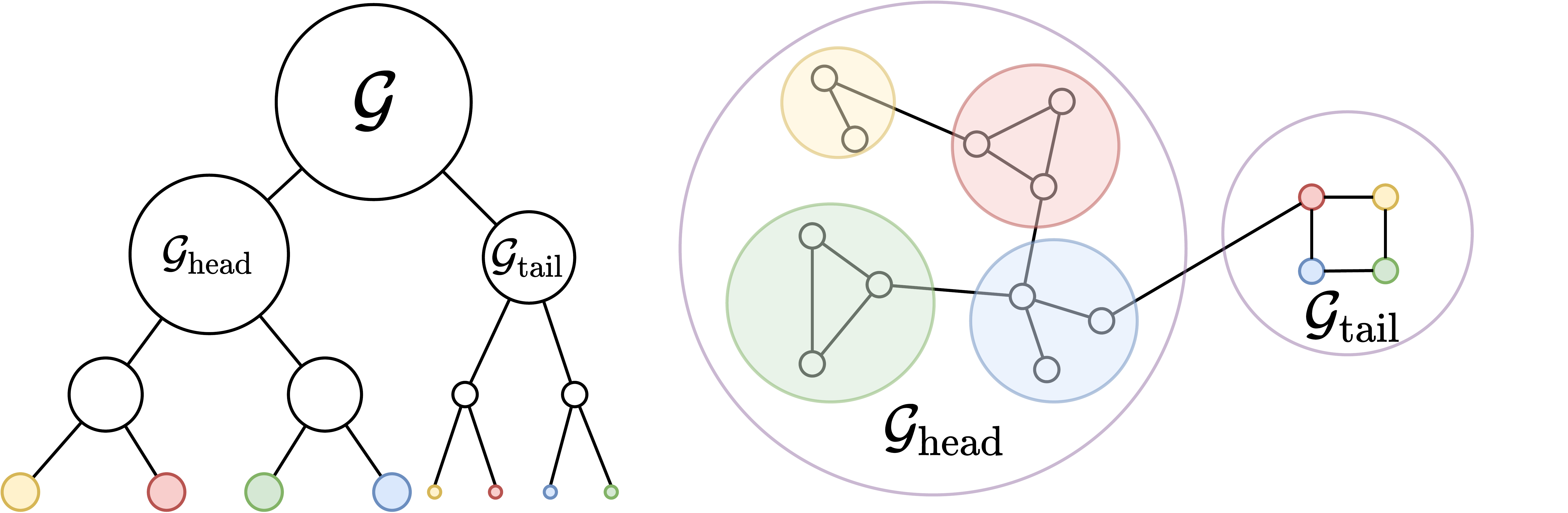}
\caption{ 
 (Top:) Pictorial representation of a quasi $3$ dimensional tensor network corresponding to a quantum circuit. The leftmost layer represents the initial state, and the rightmost layer represents the end state, where the blue circles represent measured (closed) qubits, which fix the entries in the final bitstring $\s$, and the red circles represent open qubits, where the corresponding entries in $\s$ can vary. The yellow plane $\mathcal C$ cuts into the tensor network and separates the network into two parts, $\mathcal G_{\head}$ and $\mathcal G_{tail}$, which are depicted in the left panels. The $\mathcal G_{head}$ contains all closed qubits, and $\mathcal G_{tail}$ includes all the open qubits. Both $\mathcal G_{\head}$ and $\mathcal G_{\tail}$ are further partitioned into two sub-graphs hierarchically, until size of the sub-graph is smaller than $60$. 
The bottom right panel displays the bottleneck in between $\ghead$ and $\gtail$, given by $\mathcal C$.
%group all tensors into two non-overlapping groups, $\mathcal{G}_{root}$ and $\mathcal {G}_{leaf}$. 
\label{fig:qctn}}
\end{figure}

As depicted in~Fig.~\ref{fig:qctn}, we separate $n$ qubits in the end-state (at the right-most layer in the figure) into two groups with group sizes $n_1$ (blue) and $n_2$ (red) respectively. An arbitrary bitstring $\s$ is then represented as a concatenation of partial bitstrings $\s_1$ and $\s_2$, and the probability for the bitstring can be expressed as $P_U(\s)  = P_U(\s_1;\s_2)$, the joint probability of $\s_1$ and $\s_2$. With $n_2=0$, the method is identical to the single amplitude estimation approach using tensor networks. With $n_2$ small, it is essentially the idea which has been explored in~\cite{huang2020classical,markov2018quantum,arute2019quantum}, where a number of qubits at the end state is selected manually and kept open, giving a small batch (typically $64$) of amplitudes by a single contraction. We refer to this method as \textit{batch approach}. However, there are several difficulties for the batch approach to use a large $n_2$. First, using more open qubits significantly increases the contraction complexity over single-amplitude contraction. In the literatures, only a small number (typically 6) of open qubits are selected in order to control the complexity increase. Second, finding an optimal combination of $n_2$ qubits is a hard combinatorial optimization problem, particularly when the cost for evaluating the quality of the combinations is very high. 

\paragraph{Big-head simulation---}
In this work, we aim to select a large number of open qubits $n_2$, without significantly increase the overall computational complexity. To this end, rather than selecting open qubits manually, we first find a (special) contraction order $\mathcal O$ (which indicates a sequence of pairwise contractions to perform) with a good space and time complexity, then choose the open qubits based on the order. 
%The contraction order indicates a sequence of pairwise contractions to perform. The time complexity is determined by the summation of time complexity in each pairwise contraction step, and the space complexity is defined as the size of the largest tensor encountered in each pairwise contraction.

For difficult circuits, e.g. the Sycamore circuits, the space complexity for contracting the full tensor network with a large $n_2$ open indices are typically out of reach even with a near-optimal contraction order. So we proceed by enumerating all possible partial bitstrings $\s_2$, that is, trading off space complexity using time complexity by enumerating all possible ways for closing the open qubits. However this approach causes a serious issue on the time complexity---one needs to repeat the contraction for $2^{n_2}$ times, which is intractable. In order to resolve this issue, we demand the contraction order found in the first place to satisfy a special constraint---detecting a \textit{big-head structure} of the tensor network by identifying a bottleneck, as illustrated in Fig.~\ref{fig:qctn}. 
Given an end-state bitstring $\s$, the identified bottleneck separates the whole network into two parts, the head network $\mathcal G_{\head}$ containing all the closed end-qubits, and the tail tensor network $\mathcal G_{\tail}$ containing all the open end-qubits, connected by $n_c$ edges. 
In the contraction order $\mathcal O$, $\ghead$ and $\gtail$ are contracted independently, resulting to two vectors $\vv_{\head}(\s_1)$ and $\vv_{\tail}(\s_2)$. The amplitude of $\s$ is computed at the final step of contraction in $\mathcal O$, as inner product of the two vectors $\psi(\s)=\vv_{\head}(\s_1)\cdot \vv_{\tail}(\s_2).$ Since all $n_2$ open qubits are located at $\mathcal G_{\tail}$, by fixing the partial bitstring $\s_1$, one only needs to compute $\vv_\head(\s_1)$ once, then reuses it to obtain amplitude $\psi(\s)$ for every $\s_2$. By carefully designing the contraction order $\mathcal O$, we can make the contraction cost dominated by contracting $\mathcal G_{\head}$, with the bottleneck $n_c$ small enough such that $\vv_1(\s_{\head})$ can be stored and reused. In this way, amplitudes of $2^{n_2}$ correlated bitstrings can be computed with computational complexity almost identical to that of computing one bitstring.

%\begin{figure*}[h]
%\centering
%\includegraphics[width=1.8\columnwidth]{illustrate1.pdf}
%\caption{ 
%group all tensors into two non-overlapping groups, $\mathcal{G}_{root}$ and $\mathcal {G}_{leaf}$. 
%\label{fig:qctn}}
%\end{figure*}
%\section{\label{sec:algorithm} Algorithm--}
%\begin{figure}[h]
%\centering
%%\includegraphics[width=0.9\columnwidth]{illustrate1.pdf}
%\includegraphics[width=1.0\columnwidth]{tree.pdf}
%\caption{ 
%Contraction tree and big head shape.
%%group all tensors into two non-overlapping groups, $\mathcal{G}_{root}$ and $\mathcal {G}_{leaf}$. 
%\label{fig:qctn}}
%\end{figure}
\paragraph{The algorithm---}
The key component of the big head simulation is finding a contraction order $\mathcal {O}$ satisfying the constraints described in the last section. We design the order-finding algorithm relying on a partition algorithm which splits the whole tensor network into two parts, $\mathcal G_{\head}$ and $\mathcal G_{\tail}$, using the \textit{first cut} $\mathcal C$ cutting $n_c$ edges. In Fig.~\ref{fig:qctn} we give a pictorial illustration of the first cut made on a 3-dimensional tensor network. The partition algorithm tries to minimize the cut size $n_c$, given constraints on the group sizes $\{n_1,n_2\}$.
To find a first cut and the top partition (as illustrated in Fig.~\ref{fig:qctn}), we need to make sure the computational complexity of contracting two networks is acceptable. We achieve this by hierarchically partitioning~\cite{kourtis2019fast,gray2020hyper,huang2020classical} each sub-graph into two smaller sub-graphs using a clustering algorithm until every sub-graph is small enough (set to $60$ tensors in this work). The constraint for the hierarchical partitioning is that both the time and space complexity of the individual contraction must be smaller than target values. The hierarchical partitioning also gives a contraction order $O_{\textrm{coarse}}$ to the coarse-grained graph treating the finest sub-graphs as vertices. After the partitioning, we contract all the finest sub-graphs using a greedy contraction order, then contract the coarse-grained graph according to $O_{\textrm{coarse}}$. 

For quantum circuits which are hard to simulate, e.g. the Sycamore circuits with $20$ cycles, the space complexity of the found order are still too large to store in the memory. In order to solve this problem, we employ the dynamic slicing method~\cite{chen2018classical,villalonga2019flexible,zhang2019alibaba,gray2020hyper,huang2020classical,schutski2020simple} which selects a set of $n_e$ edges from the tensor network and enumerates the indices associated with the edges. This breaks the overall contraction task into $2^{n_e}$ subtasks, each of which has much smaller space and time complexity hence can be contracted independently.

\section{\label{sec:results} Simulation of the Sycamore circuits}
We focus on Google's Sycamore circuits with $m=20$ cycles which have been used for demonstrating quantum supremacy. The sycamore circuits have $n=53$ qubits locating on a two-dimensional layout (see~Fig.\ref{fig:qctn}). Each cycle of operations contains a layer of single-qubit gates  (randomly sampled from \{$\sqrt{X},\sqrt{Y}, \sqrt{W}$\}) and two-qubit FSIM gates with different parameters $\phi$ and $\theta$. The FSIM gates have decompositional rank $4$ hence are believed to be much harder to simulate than the Controlled-Z gates~\cite{zhou2020limits} even in the approximation level. In the supremacy circuits, different pairing patterns are alternated (termed as `ABCDCDAB')~\cite{arute2019quantum}.
The circuit files were downloaded from~\cite{data}.

%20 cycles: 381 nodes，head 345, tail 36
%14 cycles: 246 nodes，head 211, tail 35

\paragraph{Computing bitstring probabilities---}
We first simplify the quantum circuit by absorbing single-qubit gates into two-qubit gates, resulting in a tensor network with $n=381$ nodes, then find a contraction order for the network which partitions the tensors into two sub-graphs ($\ghead$ and $\gtail$) with sizes $n_{\head}=345$, and $n_{\tail}=36$ respectively. We determined $21$ open qubits in $\gtail$. Without loss of generality, we assign $32$ entries of $\s_1$, i.e. the rest part of the bitstring corresponding to closed qubits, to $0$.
In contracting $\mathcal {G}_{\head}$ for obtaining the $\vv_{\head}$ vector, we use the dynamic slicing method~\cite{chen2018classical,villalonga2019flexible,zhang2019alibaba,gray2020hyper,huang2020classical,schutski2020simple} and divide the $\mathcal {G}_{\head}$-contraction task into $2^{23}$ sub-tasks, each of which has space complexity $2^{30}$ to fit into $32$G memory of GPU. 
The time complexity of contractions are listed in  Tab.~\ref{tab:complexity}.
From the table, we can see that the time complexity of contracting $\mathcal G_{\head}$ dominates $T_{2^{21}}$, the overall computational complexity for obtaining $2^{21}$ bitstring amplitudes. 
The computational cost for obtaining a different number of bitstrings in different algorithms is compared in~Tab.\ref{tab:compare}. We remark that our computational cost for obtaining $2$ million bitstring probabilities is much lower than the estimated computational cost of the Shr\"odinger-Feynman method used by Google~\cite{arute2019quantum}, and is also lower than the computational complexity for obtaining $64$ amplitudes in the state-of-the-art tensor network method~\cite{huang2020classical}. Our algorithm can be trivially parallelized on multiple GPUs. So in our experiments, we employed a small computational cluster composed of $48$ NVIDIA V100 and $12$ A100 GPUs. We ran all the sub-tasks independently on GPUs (sharing with other user jobs), which are trivially parallelized without any communications among them. The overall computation cost about $5$ days. More details about the algorithm and the computations can be found in the Appendices.
\begin{table}[]
\begin{tabular}{|c|c|c|c|c|c|}
    \hline
    \#subtasks & $S_{\text{total}}$ & $T_{\text{sub}}$  & $T_{\text{head}}$  & $T_{\text{tail}}$ & $T_{\text{total}}$  \\
    \hline
    $2^{23}$& $2^{30}$ & $5.37\times 10^{11}$  & $4.51\times 10^{18}$  & $2.87\times 10^{15}$&$4.51\times 10^{18}$\\
\hline
\end{tabular}
\caption{The number of sub-tasks (\#subtasks), total space complexity $S_{total}$, time complexity for a single subtask $T_{sub}$, time complexity for contracting the head network,  $T_{\head}$, time complexity for contracting the tail network $T_{\tail}$, and time complexity for obtaining totally $2^{21}$ bitstrings $T_{\tail}$ in the simulation of Sycamore circuit with $20$ cycles.\label{tab:complexity}}
\end{table}

\begin{table*}[htb]
\begin{tabular}{|c|c|c|c|c|c|c|}
\hline
&\# bitstrings &Time complexity & Space complexity &  Computational time & Computational hardware\\
\hline
Google~\cite{arute2019quantum}& $10^6$ & ---&---  & 10,000 years &Summit supercomputer\\
\hline
Cotengra~\cite{gray2020hyper}& 1 &$3.10 \times 10^{22}$ & $2^{27}$  & 3,088 years &  One NVIDIA Quadro P2000 \\
\hline
Alibaba~\cite{huang2020classical} &64& $6.66\times10^{18}$ & $2^{29}$  & 267 days & One V100 GPU\\
\hline
Ours & $\mathbf{2097152}$ &$4.51\times 10^{18}$ & $2^{30}$ & 149 Days & One A100 GPU\\
\hline
\end{tabular}
\caption{Comparison of computational cost among different methods on Sycamore circuit with $53$ qubits and $20$ cycles. \label{tab:compare}}
\end{table*}

\paragraph{Fidelity---}
A standard test for accuracy of obtained $L$ bitstrings, $\{\s_i \} _{i=1}^{L}$, is the linear cross entropy benchmark (XEB) fidelity~\cite{aaronson2016complexity,boixo2018characterizing,arute2019quantum},
\begin{equation}\label{eq:xeb}
F_{\mathrm{XEB}} = \frac{2^n}{L}\sum_{i=1}^L P_U(\s_i)-1.
\end{equation}
Notice that $P_U(\s_i)$ is the probability of the bitstring in the circuit $U$, which requires an exact evaluation of the amplitudes.
When the samples are generated from the Porter-Thomas distribution~\cite{porter1956fluctuations,brody1981random,boixo2018characterizing}
\begin{equation}\label{eq:pt}
%\mathrm{Prob}(p) = 2^ne^{-p2^n},
\mathrm{Prob}(p) = 2^n\exp(-p2^n),
\end{equation}
one has $\xeb=1$. On the other hand, if the bitstrings are sampled from the  random uniform distribution, one would have $F_{\textrm{XEB}}=0$. In Google's quantum supremacy experiments, they achieved $F_{\textrm{XEB}}=0.002$.

\begin{figure}[h]
\centering
\includegraphics[width=0.5\columnwidth]{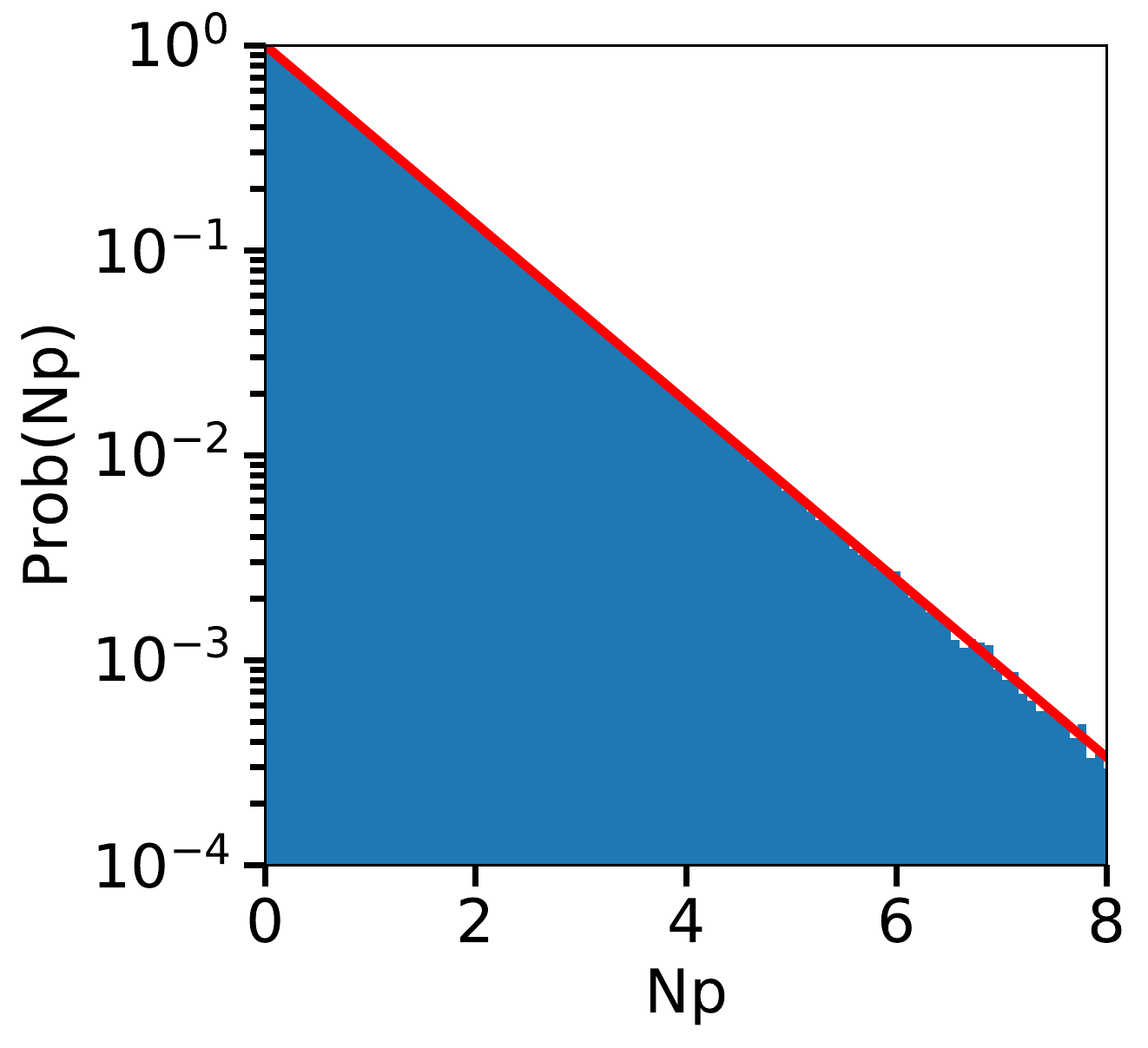}
\includegraphics[width=0.49\columnwidth]{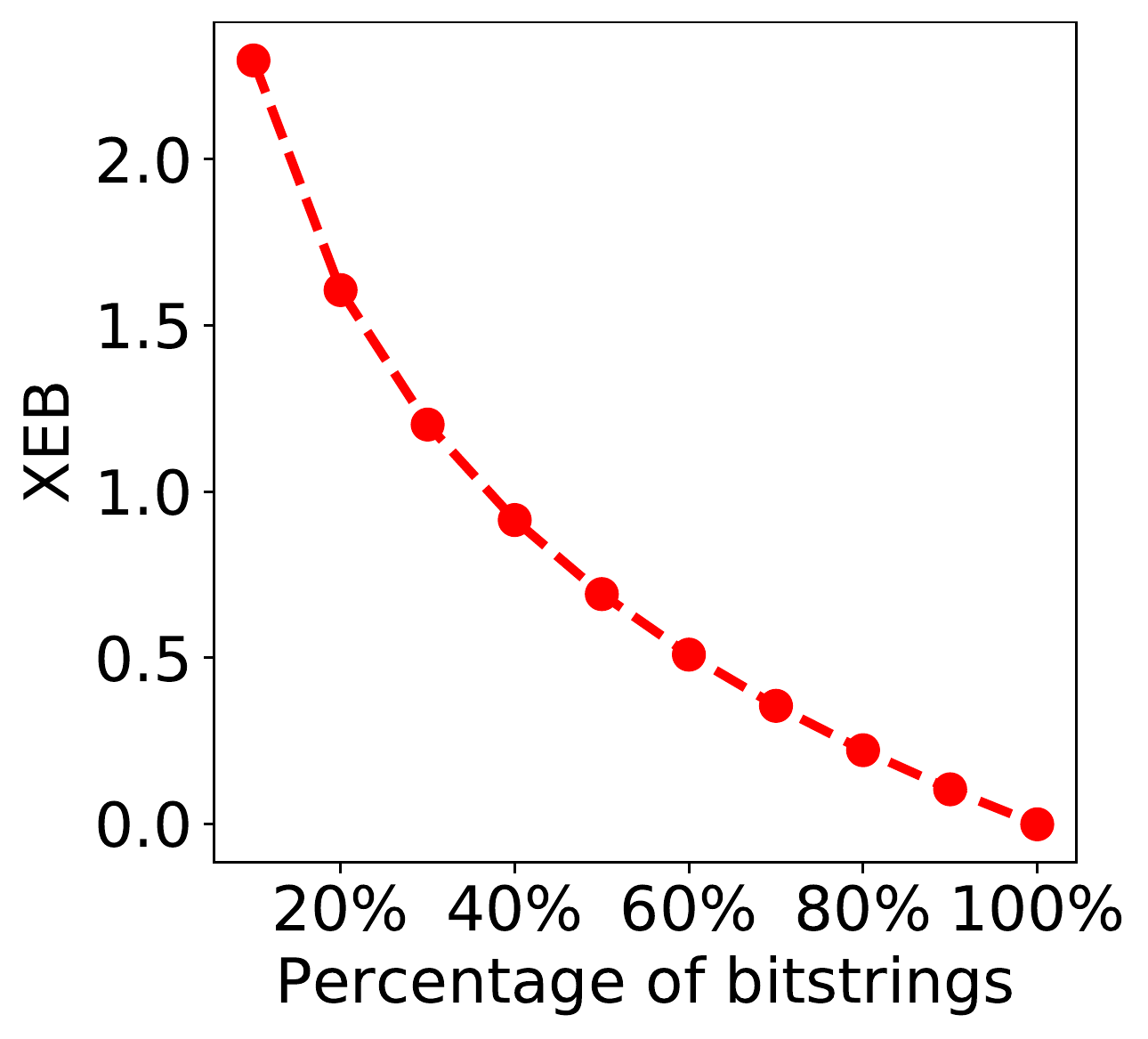}
\caption{ 
(Left): Histogram of bitstring probabilities  $P_U(\s) = P_U(\s_1;\s_2)$ for $L=2^{21}$ bitstrings obtained from the Sycamore circuit with $n=53$ qubits, $m=20$ cycles, sequence ABCDCDAB, seed $0$, and the assignment of partial bitstring $\s_1$ are fixed to 
$\underbrace{0,0,0,\cdots,0}_{32}$. 
In the figure, $N=2^{53}$, $p$ denotes the probability of bitstring $P_U(\s)$, and the red line represents the Porter-Thomas distribution~\eqref{eq:pt}. The $\xeb$ for all the bitstrings is $-0.000926$.
(Right): $\xeb$ computed for an ensemble of bitstrings which are post-selected from total $2^{21}$ bitstrings sorted by probabilities.
\label{fig:dist}}
\end{figure}

In this work, without loss of generality, we fix $32$ entries as  $\s_1 = \underbrace{0,0,0,\cdots,0}_{32}$, and enumerate all possible combinations of other $21$ entries in the bitstring. This produces a set of $2^{21}$ correlated bitstrings, denoted as 
%$\Omega = \underbrace{0,0,0,\cdots,0}_{32} \otimes \{0,1\} ^{21}\in \{0,1\} ^{53}$. 
$\Omega = \{\s_1;\s_2^{(i)} \}_{i=1}^{L}$, 
where $L=2^{21}$.
Some samples of bitstrings can be found in Appendices. Since we assign equal weight for all $2^{21}$ bitstrings, the distribution of the bitstrings can be regarded as uniform distribution $P_{\textrm{gen}}(\s_1;\s_2)=2^{-21}$ for all $2^{21}$ bitstrings, given an assignment of the rest $32$ entries $\s_1$ which the $\xeb$ should depend on. 
We plot the histogram of the obtained $2^{21}$ bitstrings in Fig.~\ref{fig:dist}, where we can see that the obtained distribution is very close to the Porter-Thomas distribution, with $\xeb$ value equals to $-0.000926$, which is close to $0$. The minimum and the maximum probability of bitstrings are $P_{\textrm{min}}=8.04\times 10^{-8}\times 2^{-53}\approx 0$ and $P_{\textrm{max}}=16.1\times 2^{-53}$ respectively.
Thus, we are able to post-select a subset of bitstrings such that the XEB of the selected bitstrings varies in between $\xeb^{\textrm{min}}=-1.0$ and $\xeb^{\textrm{max}}=15.1$. In Fig.~\ref{fig:dist} we sort the bitstring probabilities then plot the $\xeb$ value as a function of fraction of bitstrings in the post-selection. In particular, by selecting bitstrings with top probabilities, we arrive at $10^6$ bitstrings with $\xeb=0.739$.
We remark that for reaching the $\xeb$, the probability for all bitstrings are computed exactly. This is in contrast to the previous methods which require a sampling method, e.g. the \textit{frugal sampling}~\cite{markov2018quantum,arute2019quantum}, to mix a small number of bitstrings with (known) large probabilities to a large number of random bitstrings with unknown probabilities. 

%$5290$ hours total GPU times.
%5738 hours running time.

%In practice, we observed that the computation time for contracting a sub-task is $1.2$ seconds on an NVIDIA A100 GPU with $80$G memory, and is $2.9$ seconds on a V100 GPU with $32$G memory.
%The time complexity for each sub-task is 
%%$2.088\times 10^{12}$. 
%$5.37\times 10^{11}$, and the total time complexity for contracting $\mathcal G_{head}$ is $4.51\times 10^{18}$.
%As compared to the contraction of $\mathcal G_{head}$, the cost of contracting $\mathcal G_{tail}$ is very light, with total time complexity being $2.87\times 10^{15}$.
%

%$1.347\times 10^9$
%$1.75\times 10^{19}$
%$10^{12.319631274396317}$, 
%space complexity is $2^{30}$, it takes $1.235$ seconds on an NVIDIA A100 GPU with 40G memory, and takes $2.919$ seconds on a NVIDIA V100 GPU with 32G memory. The efficiency (see appendix) for A100 and V100 is $69.4\%$, $40.5\%$ respectively. On the open qubits part, there are $21$ open qubits, time complexity is 
%$1.347\times 10^9$
%$10^{9.129287319592848}$ $
%and space complexity is $2^{23}$.

% For Sycamore circuit of $53$ qubits and $14$ cycles with ABCDCDAB sequence, $9$ edges are sliced so there are overall $2^{9}$ sub-tasks. For each sub-task, time complexity is $2.543\times 10^{11}$,
% %$10^{12.319631274396317}$, 
% space complexity is $2^{30}$, it takes $0.541$ seconds on a V100S(32G memory, 16.35TFLOPS) to complete a sub-task
% and the GPU efficiency is $23.02\%$. On the open qubits part, there are $19$ open qubits, time complexity is 
% $1.893\times 10^7$ and space complexity is $2^{21}$.

\section{Discussions}

We have presented the big-head tensor-network method for computing a large number of bitstring probabilities for quantum circuits. The goal of Google's quantum supremacy experiments was to obtain a large number of samples achieving a high enough $\xeb$ for the Sycamore circuits with sufficient depth such that the task is intractable for classical computing. We have demonstrated that with our algorithm we are able to classically obtain a large number of samples with an even larger $\xeb$ than the Sycamore experiments.

Our algorithm has several advantages over Google's hardware sampling of the Sycamore circuits.  
First, our method is able to output the exact amplitude and probability of any bitstring, while Google's samples are not verified, with XEB estimated using extrapolations; second, our method is less noisy than Google's experiments in obtaining samples from the Sycamore circuits, because the XEB of the post-selected $1$ million bitstrings obtained using our method is higher than those in Google's experiments; third, we are able to compute conditional probabilities $P_U(\mathbf s_2|\mathbf s_1)$ and sample from this distribution accordingly, which is hard for Google's quantum circuit hardwares.
At the same time, our experiments also reflect that Google's hardware has several advantages over our algorithm.
The most significant one is that Google's hardware is much faster in sampling the quantum circuits with sufficient depth, while our algorithm has exponential complexity, hence is not scalable to both depth and qubit number; Although being noisy, Googles' samples are not correlated to each other, while in this work we generated correlated samples, which belong to a subspace of the whole bitstring space~\footnote{Notice that the correlation issue of our method can be significantly relaxed by selecting some large-probability bitstrings, then mixing them with a large number of random samples. For example, if we select $220$ bitstrings with top probabilities and mix with $999780$ random bitstrings, the correlations among the overall $1$ million bitstrings would be negligible, while the $\xeb$ would be $0.002$, the same as Google's experiments.}.
As a conclusion, we would like to express our opinion in a constructive way that based on the newly developed tensor network simulation methods, it would be very interesting to combine classical computations and NISQ quantum computations, for solving challenging problems in the real-world.

Although we have focused on the simulation of Google's Sycamore circuits, our algorithm is generally designed. The big-head shape and bottleneck structure are general phenomenons existing in many tensor networks, as well as in many real-world networks~\cite{Zhang2014pnas}, which can be detected using clustering algorithms~\cite{Krzakala2013}. The proposed algorithm can be used straightforwardly for simulating and verifying existing and near-future NISQ quantum circuits.
%We believe that based on our algorithm, by exploring better order-finding algorithms, and improving computational efficiency, it would possible to compute more bitstring probabilities more efficiently.
Our code is implemented straightforwardly using \textit{Python} and \textit{Pytorch}. The contraction code together with the contraction orders, slicing indices, and the obtained bitstrings probabilities, are available at ~\cite{code}.

\paragraph{Acknowledgements--}
%\begin{acknowledgments}
We thank Xun Gao, Fengyao Hou, Jinguo Liu, Dingshun Lv, Yibo Yang, and Lei Wang for helpful discussions. We are grateful to Haijun Liao and Lei Wang for offering A100 GPUs. 
%P.Z. is supported by Key Research Program of Frontier Sciences, CAS, Grant No. QYZDB-SSW-SYS032, and Project 12047503 and 11975294 of National Natural Science Foundation of China. Part of the computation was carried out at the High-Performance Computational Cluster of ITP, CAS.
%\end{acknowledgments}

%\bibliographystyle{apsrev4-1}
\bibliography{refs.bib}% Produces the bibliography via BibTex

\clearpage

\onecolumngrid
\appendix  
\balance

\section{Details of the contraction algorithm}
Based on the big-head simulation scheme, our contraction algorithm is mainly composed of two parts, the hierarchical partitioning based algorithm for finding the first cut and a contraction order with tractable computational cost, and the dynamic slicing which breaks the overall giant contraction process into many small and tractable contraction processes, each of which can be contracted independently and parallelly.

\paragraph{Finding contraction order using hierarchical partitioning}

An essential problem in tensor network contraction is to determine an order for contracting tensors pair by pair, which is referred to as \textit{contraction order}. For tensor networks defined on $1$D or $2$D lattices the optimal order can be easily found. However, the tensor networks associated with quantum circuits are usually irregular, it is a difficult problem to find an optimal order for contracting the whole network. Lots of effort has been devoted in finding a good contraction order. Markov and Shi~\cite{markov2008simulating} first showed that one can use the optimal tree decomposition of the line graph associated with the tensor network to find the optimal contraction order. 
However, finding the optimal tree decomposition for a general graph is NP-hard problem, thus one usually adopts heuristic algorithms, such as the branch-and-bound algorithm with a restricted running time, to find a good tree decomposition. 

In~\cite{kourtis2019fast}, the authors proposed to use graph partitioning algorithms to find a good contraction order for contracting tensor networks associated with constraint satisfaction problems. Later in ~\cite{gray2020hyper}, the method was further developed and applied to the quantum circuit simulations. In~\cite{huang2020classical}, further improvements were made, and authors were able to justify that contracting the Sycamore circuits with $20$ circles is reachable using a supercomputer.

In this work, the order-finding algorithm is responsible not only for low computational cost, but also for detecting the big-head structure of the network, so that the $\head$ network is only needed to contracted once and reused for computing all the bitstring probabilities defined by enumerating the assignment of $\s_2$ in $\gtail$.

\begin{figure}[htb]
\centering
\includegraphics[width=0.5\columnwidth]{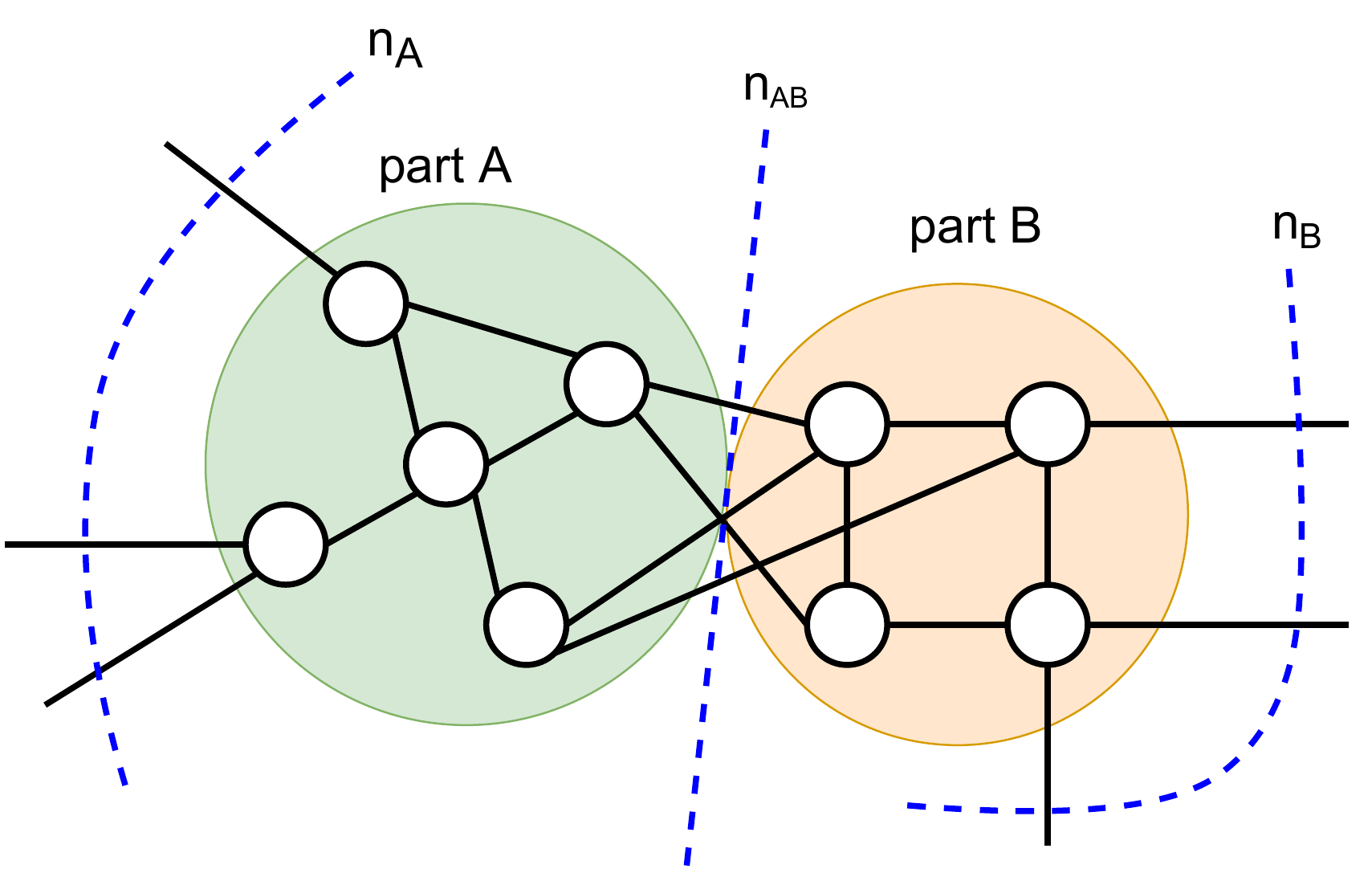}
\caption{Illustration of the bipartition. $n_{AB}$ is the number of indices shared by sub-graphs A and B, with $n_A$ and $n_B$ representing the number of outer indices owned by two sub-graphs respectively. The bond dimension of each edge studied in network associated with the quantum circuit is $2$, the time complexity of the contraction is $2^{n_A + n_B + n_{AB}}$ 
\label{fig:bipartition}}
\end{figure}

The most important step of our approach is the first cut, given by the initial partition, which separates the tensor network corresponding to quantum circuits into the head and tail tensor networks. Notice that we set a constraint to the first cut such that the pair contraction associated with subsequent bi-partitions will never create time and space complexity larger than the target values. If the first cut does not satisfy this constraint, we will try other partitions until the constraints are satisfied.
Based on the first cut, we use a hierarchical bipartition algorithm, which sequentially and hierarchically divides the remaining tensor network into two parts. Each bipartition represents a binary structure in the contraction tree, thus when the bipartition process ends (if the number of tensors in the sub-graph is smaller than $60$), it provides a contraction order. If we carefully restrict the complexity of each bipartition, the time and space complexity of the resulting order will also be restricted to the corresponding initial partition can be validated. The complexity of one bipartition is depicted in Fig~\ref{fig:bipartition}.

\paragraph{Dynamic slicing}
For large quantum circuits with a sufficient depth, even with a very good contraction order, the space complexity is usually out of reach, beyond the capacity of RAM and GPU memory.
For example, the largest tensor we encountered in contracting the Sycamore circuits with $n=53$ qubits and $m=20$ cycles has an overall size of $2^{53}$ . If we use single-precision complex numbers, it will take roughly $13421$ TB memory to store the giant tensor. In order to make contraction possible using current computational hardwares, we need techniques to separate the whole contraction task into many sub-tasks, each of which requires much less memory than the original contraction. 
Slicing is a technique of removing some of the indices in the tensor network, enumerating their dimensions as subtasks, and sum over subtasks to get an equivalent result. Dynamic slicing~\cite{chen2018classical} or sliced subtree reconfiguration~\cite{gray2020hyper} improve the vanilla slicing method by fine-tuning the contraction tree each time an index being sliced, this gives a much smaller slicing overhead (proportion of time complexity after and before slicing).
In our experiments for contracting the $\mathcal {G}_{\head}$ to obtain the $\vv_{\head}$ vector, using the dynamic slicing we divided the $\mathcal {G}_{\head}$-contraction task into $2^{23}$ sub-tasks, each of which has space complexity $2^{30}$ to fit into $32$G memory of NVIDIA V100 GPU.

% For 20 cycles, it took about 5290 GPU hours and 5738 CPU + GPU hours for the head tensor network simulation. While for 14 cycles, it took about 271.3 GPU seconds and 297 CPU+GPU seconds. The simulation of tail tensor network will only take less than one second which is negligible.

\begin{table*}[]
\begin{tabular}{|c|c|c|}
\hline
bitstring&amplitude&probability\\
\hline
\color{blue}00000000000\color{red}000\color{blue}00000\color{red}00001\color{blue}0000\color{red}01101\color{blue}0000\color{red}00100\color{blue}00\color{red}100\color{blue}000000 &$3.62\times 10^{-8}$ & $1.31\times 10^{-15}$ \\ % #11556
%\hline 
\color{blue}00000000000\color{red}000\color{blue}00000\color{red}00000\color{blue}0000\color{red}00000\color{blue}0000\color{red}00001\color{blue}00\color{red}001\color{blue}000000 &$1.55\times 10^{-8}$ & $2.39\times 10^{-16}$ \\ % #9
%\hline 
\color{blue}00000000000\color{red}111\color{blue}00000\color{red}11111\color{blue}0000\color{red}11111\color{blue}0000\color{red}11111\color{blue}00\color{red}110\color{blue}000000 &$6.3\times 10^{-9}$ & $3.97\times 10^{-17}$ \\ % #2097150
%\hline
\color{blue}00000000000\color{red}111\color{blue}00000\color{red}11111\color{blue}0000\color{red}11111\color{blue}0000\color{red}11110\color{blue}00\color{red}101\color{blue}000000 &$3.14\times 10^{-9}$ & $9.86\times 10^{-18}$ \\ % #2097141
%\hline
\color{blue}00000000000\color{red}000\color{blue}00000\color{red}00000\color{blue}0000\color{red}01000\color{blue}0000\color{red}00111\color{blue}00\color{red}010\color{blue}000000 &$9.17\times 10^{-10}$ & $8.41\times 10^{-19}$ \\ % # 2106
%\hline 
\hline
\end{tabular}
\caption{Examples of bitstrings with their amplitudes and probabilities computed using our algorithm for the Sycamore circuit with $53$ qubits and $20$ cycles.\label{tab:bitstrings}}
\end{table*}

%\subsection{Branch merging}

\section{\label{details} Details of simulations}
During the tensor network contractions, we used single-precision complex numbers (complex64) and the Pytorch library. As a check, we have compared the contraction results of single sub-task tensor networks to that using complex single and double precision, and also evaluated the effect of extracting the largest number or tensor norm from the input and output of each tensors contractions, and concluded that complex single computation preserves enough precision that is necessary for evaluating the amplitude of bitstrings. 

\paragraph{GPU efficiency---}
One can define the GPU efficiency as
\begin{equation}
    E = \frac{8\times\text{time complexity}}{\text{GPU FLOPS capacity}\times\text{running time}}\; .
\end{equation}
Here 8 is the factor for the FLOPS count with complex tensor operations (2 for addition and 6 for multiplication). 

Directly running the sub-tasks on GPU gives a low GPU efficiency, because a typical order for contracting $\head$ involves many tensor operations between a large tensor and a small tensor. To overcome this problem, we merge two small tensors into a larger tensor before operating with the large tensor. This approach is called branch merging as introduced in~\cite{huang2020classical}. It balances the size of tensor operations in the contraction order hence hugely increases the GPU efficiency, with only a small cost of time complexity increase. In our numerical experiments for the Sycamore circuits with $n=53$ qubits and $m=20$ cycles, it raises the GPU efficiency in V100 GPUs from $4.4\%$ to $38.4\%$ by introducing an additional $3.87$ times of time complexity. And the overall computational time will be around $55\%$ lower than the original one.
In our experiments, the GPU efficiency of %V100 (32GB memory, 14.13 TFLOPs theoretical single precision performance) is $40.49\%$, and the efficiency for 
NVIDIA A100 (40GB memory, 19.5 TFLOPs) GPU is $69.40\%$. 

\begin{figure}[h]
\centering
\includegraphics[width=0.9\columnwidth]{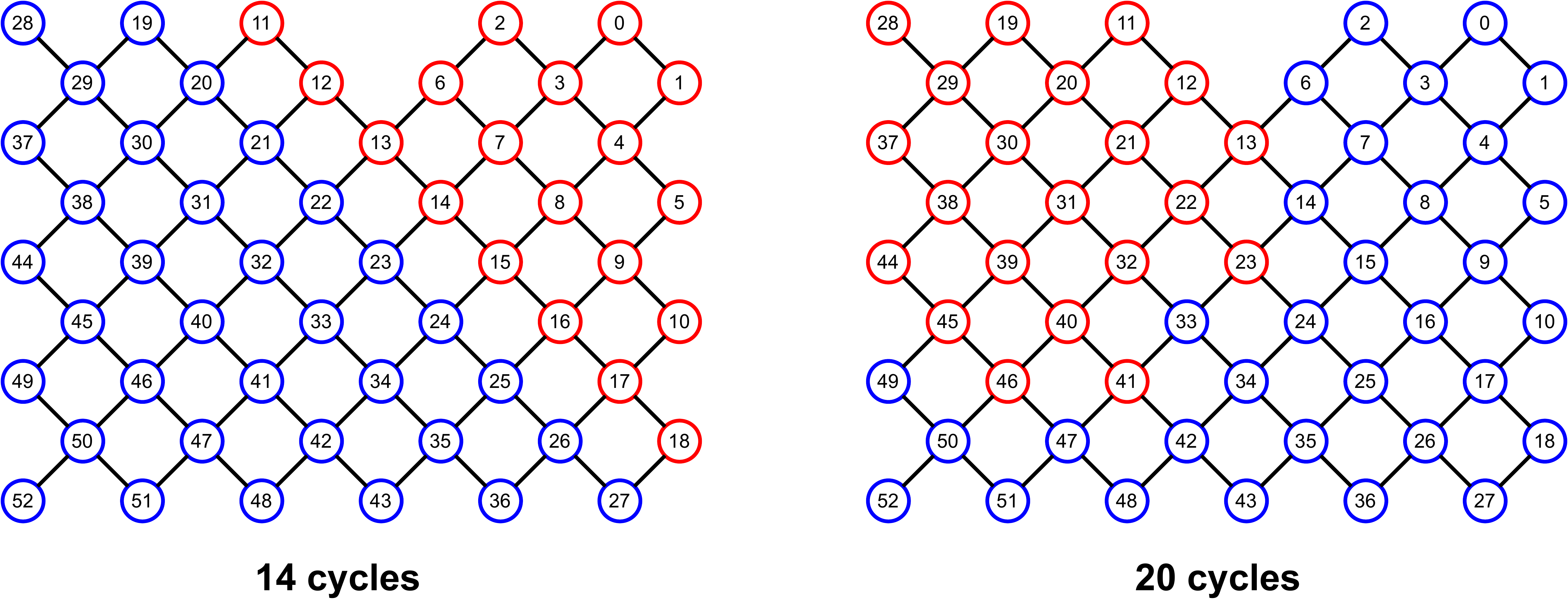}
\caption{ 
The layout of the sycamore circuits. Colors represent a division of all $53$ qubits at the end state into open (red) and close (blue) qubits. The numbering is according to that in the Google's circuit data~\cite{data}.
\label{fig:layout2}}
\end{figure}

\paragraph{Computation details---} For the Sycamore circuits with $53$ qubits, based on a contraction order we identify open and closed qubits as shown in Fig.~\ref{fig:layout2}.
For $m=20$ cycles, there are $21$ open qubits (qubit ids: 11, 12, 13, 19, 20, 21, 22, 23, 28, 29, 30, 31, 32, 37, 38, 39, 40, 41, 44, 45, 46), time complexity for contracting the tail network with an assignment of $\s_2$ is $1.347\times 10^9$ %$10^{9.129287319592848}$ $
and space complexity is $2^{23}$.  
for the $\head$ part in order to make contraction fit into the GPU memory, $23$ edges are sliced so that there are overall $2^{23}$ sub-tasks. For each sub-task, time complexity is $2.088\times 10^{12}$,
%$10^{12.319631274396317}$, 
space complexity is $2^{30}$, it takes $1.235$ seconds for contraction $1.5336$ seconds for overall calculation on an NVIDIA A100 GPU, $2.919$ seconds and $3.044$ seconds respectively on a NVIDIA V100 GPU. The overall calculation means that all the computation process of a sub-task of contraction, including tensor preparation, slicing, contraction, and summation of the final tensors. 
Some samples of obtained bitstring probabilities are listed in~Table~\ref{tab:bitstrings}.

\begin{figure}[h]
\centering
\includegraphics[width=0.46\columnwidth]{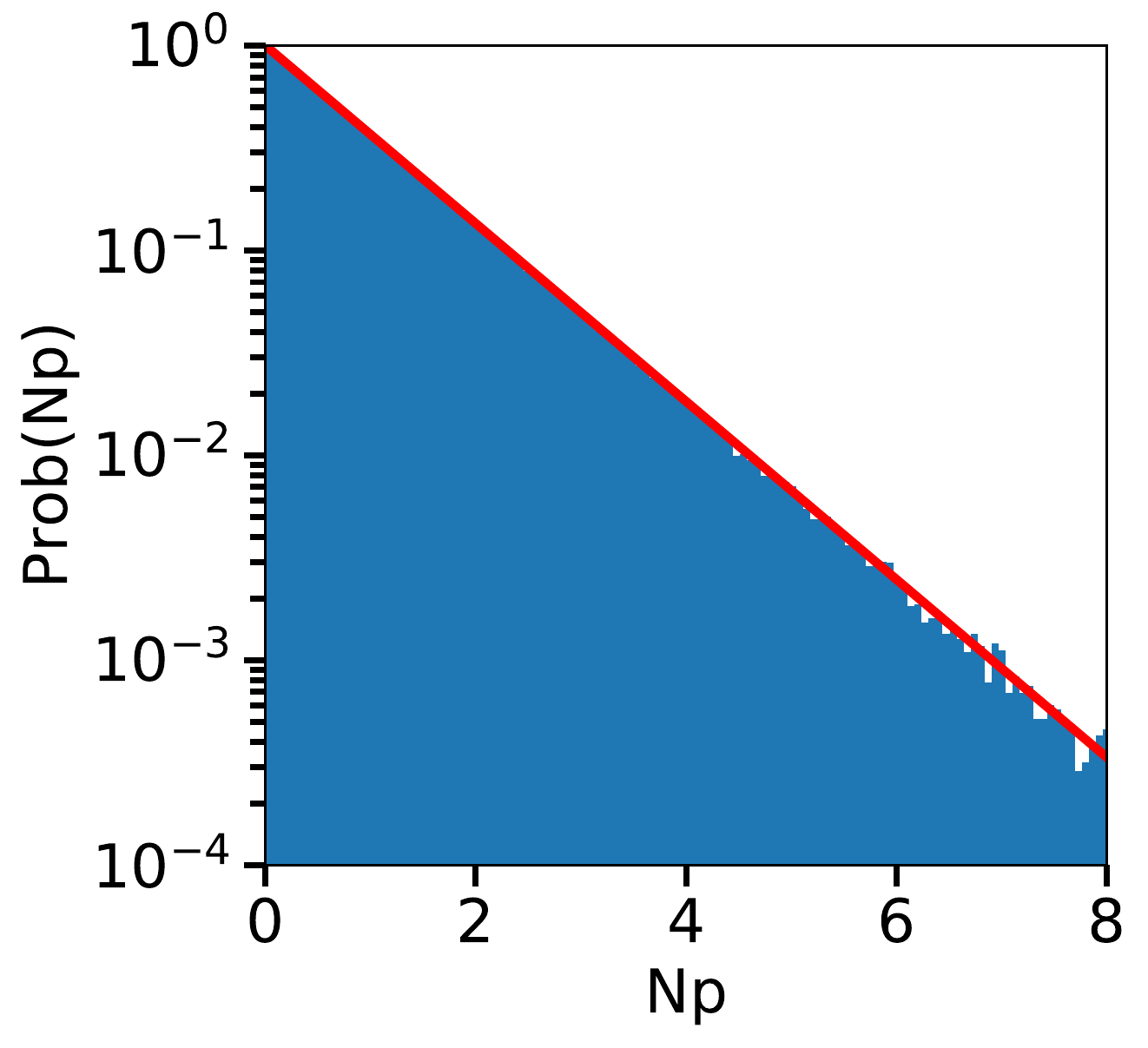}
\includegraphics[width=0.43\columnwidth]{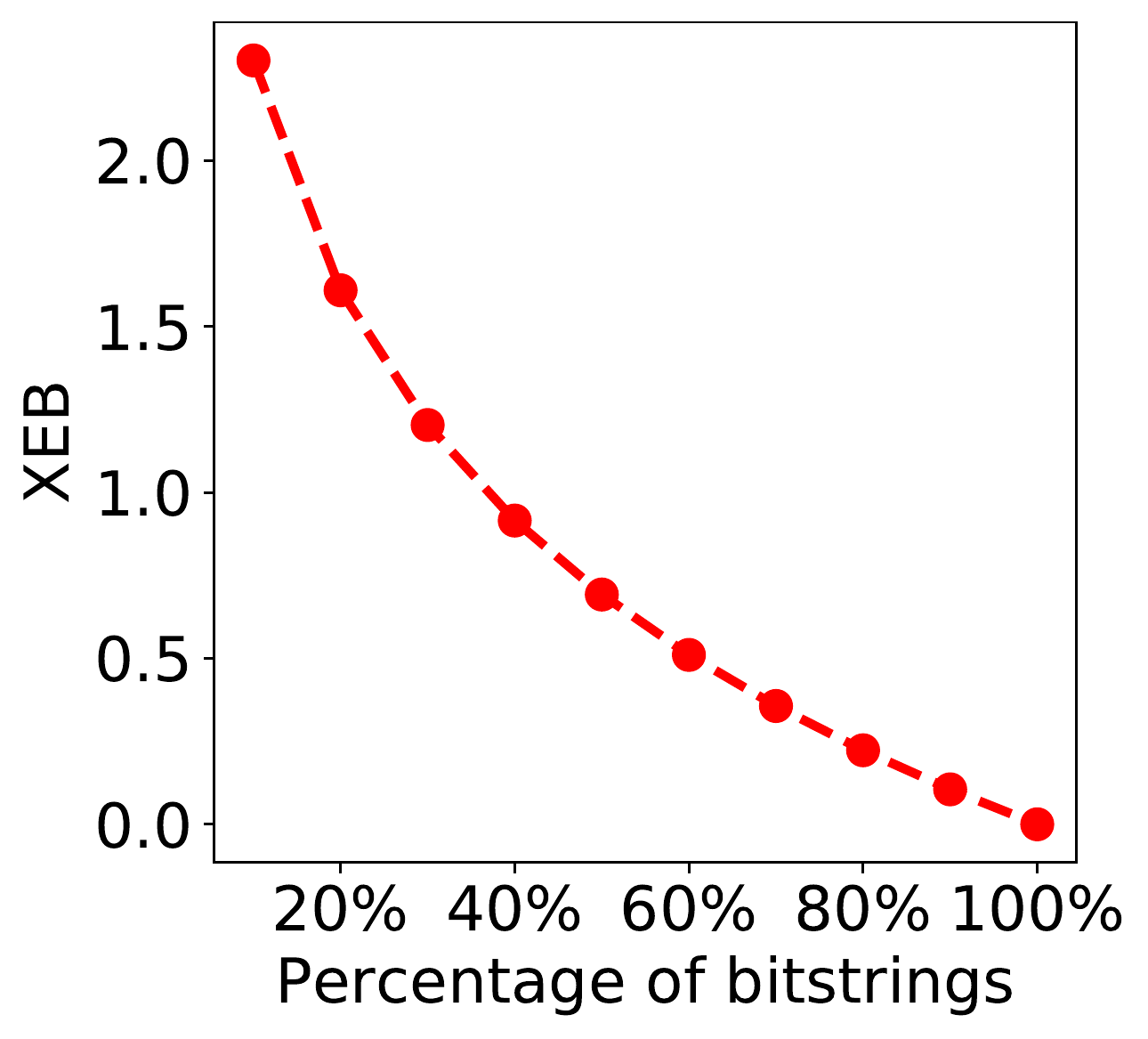}
\caption{ 
(Left): Histogram of bitstring probabilities  $P_U(\s) = P_U(\s_1;\s_2)$ for $L=2^{19}$ bitstrings obtained from the Sycamore circuit with $n=53$ qubits, $m=14$ cycles, sequence ABCDCDAB, seed $0$, and the assignment of partial bitstring $\s_1$ are fixed to 
$\underbrace{0,0,0,\cdots,0}_{34}$. 
In the figure, $N=2^{53}$, $p$ denotes the probability of bitstring $P_U(\s)$, and the red line represents the Porter-Thomas distribution~\eqref{eq:pt}. The $\xeb$ for all bit-strings is $-0.00687$.
(Right): $\xeb$ computed for an ensemble of bitstrings which are post-selected from total $2^{19}$ bitstrings sorted by probabilities.  \label{fig:dist14} }
\end{figure}

For the Sycamore circuits with $53$ qubits and $14$ cycles, ABCDCDAB sequence and seed $0$, $9$ edges are sliced so there are overall $2^{9}$ sub-tasks. For each sub-task, time complexity is $2.543\times 10^{11}$, space complexity is $2^{30}$, it takes $0.541$ seconds on a V100S (32GB memory, 16.35 TFLOPs) to complete a sub-task and the GPU efficiency is $23.02\%$. On the tail part, there are $19$ open qubits (qubit ids: 0, 1, 2, 3, 4, 5, 6, 7, 8, 9, 10, 11, 12, 13, 14, 15, 16, 17, 18), time complexity is 
$1.893\times 10^7$ and space complexity is $2^{21}$. Calculation time on the tail part of both $14$ and $20$ cycles are negligible compared to the head part.
The results for $m=14$ cycles are shown in Fig.~\ref{fig:dist14}, where we can see that the histogram of bistring probabilities deviate a little from the Porter-Thomas distribution, giving $\xeb=-0.00687$.

\section{Distribution of conditional probabilities and marginal probability}
The probability of a bitstring is the joint probability over $\s_1$ and $\s_2$, i.e. $P_U(\s)=P_U(\s_1;\s_2)$. In our settings, entries of $\s_1$ is fixed to an arbitrary assignment, and we are able to compute probabilities for all possible bitstrings obtained by enumerating $\s_2$. This means that in addition to computing $P_U(\s)$, we are also able to compute the marginal probability $$P_U(\s_1)=\sum_{\s_2}P_U(\s_1;\s_2),$$
and conditional probability distribution 
$$P_U(\s_2|\s_1)=\frac{P_U(\s_1;\s_2)}{P_U(\s_1)}.$$
Apparently, $P_U(\s_1)$ is nothing but $\xeb+1$, hence in our experiments (in the main text) for the Sycamore circuits with $n=53$ qubits and $m=20$ cycles, $P_U(\s_1) = 0.999*2^{-n_1}$, which is really close to $2^{-n_1}$. For the random quantum circuits, we anticipate that the expectation of $P_U(\s_1)$ averaged over $\s_1$ should be around $\langle P_U(\s_1)\rangle_{\s_1}=2^{-n_1}$. We see that according to  the computed $P_U(\s_1)$ value, an instance of $P_U(\s_1)$ is indeed tightly around this expectation.

\begin{figure}[h]
\centering
\includegraphics[width=0.45\columnwidth]{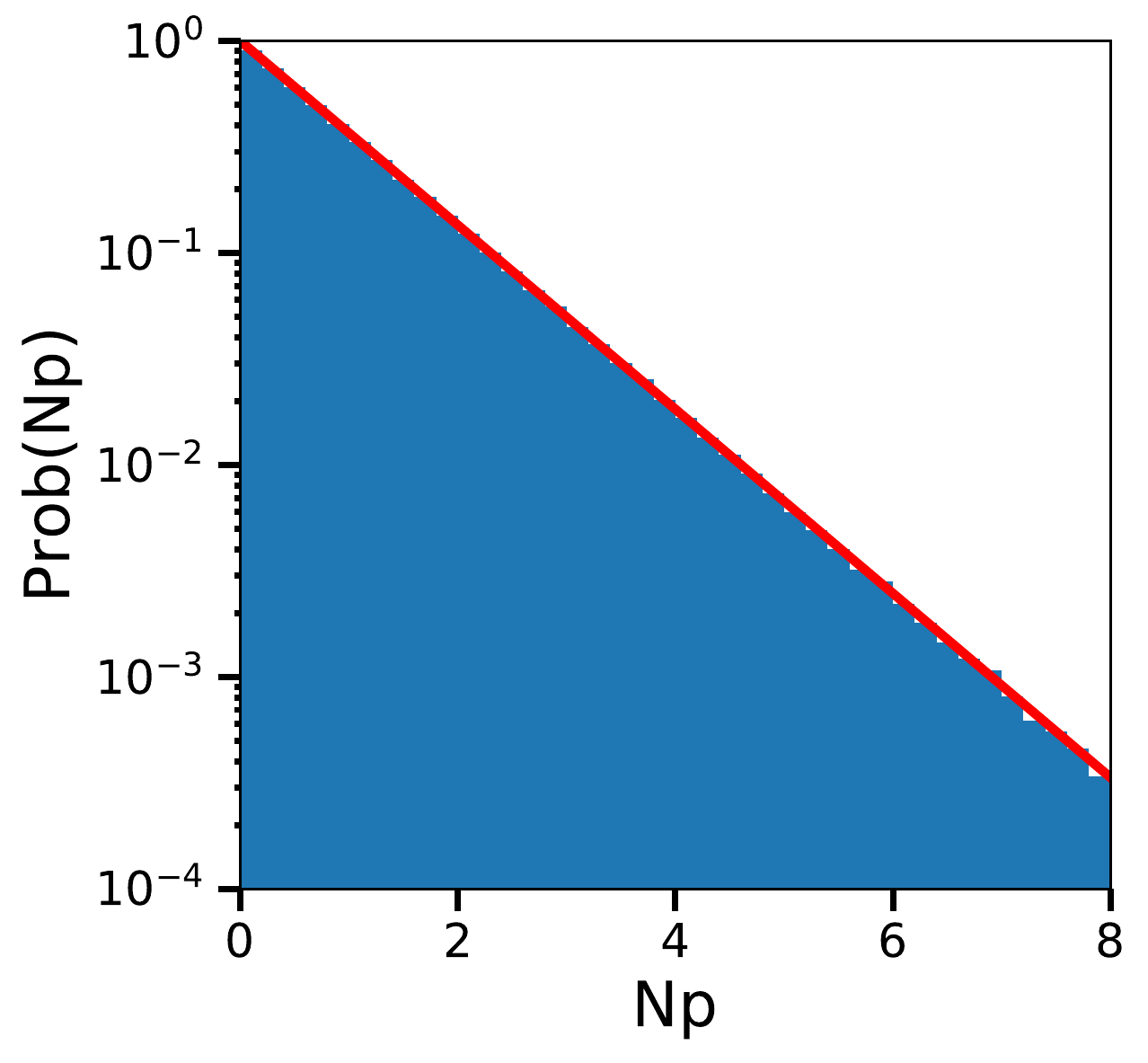}
\includegraphics[width=0.45\columnwidth]{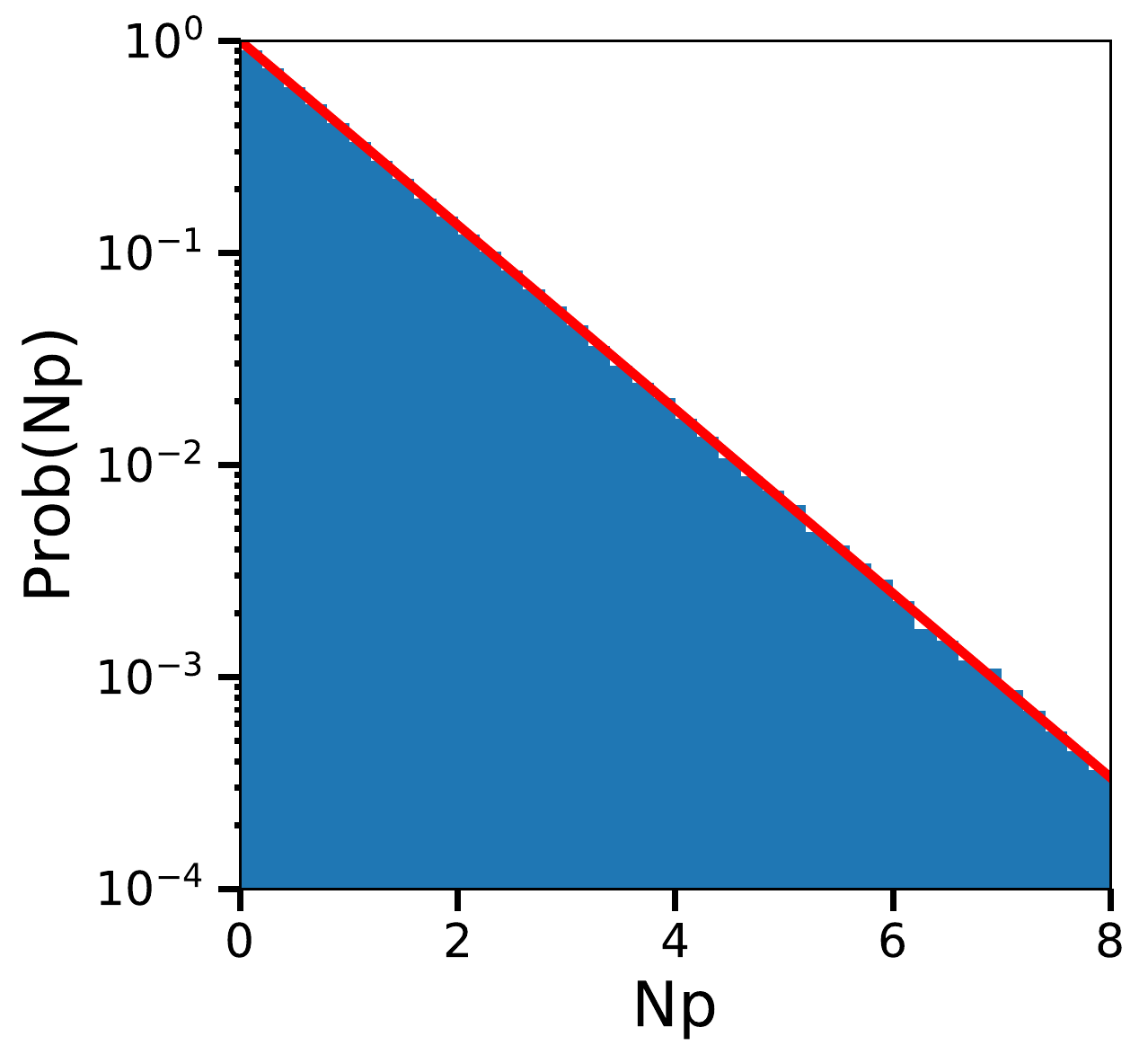}
\caption{ 
Histogram of conditional probability $P(\s_1|\s_2)$ for $L=2^{21}$ bitstrings obtained from the Sycamore circuit with $n=53$ qubits, $m=20$ cycles (Left) and $m=14$ cycles (Right), seed $0$, sequence ABCDCDAB. In the figure, $N=2^{21}$ for the left panel and $N=2^{19}$ for the right panel, $p$ denotes the conditional probability $P_U(\s_2|\s_1)$, and the red line represents the Porter-Thomas distribution~\eqref{eq:pt}. The XEB~\eqref{eq:xeb} for the conditional probability distribution is exactly $0$ in both panels. 
%$0.990686$
%$0.03167797$
\label{fig:cond}}
\end{figure}

In Fig.~\ref{fig:cond} we plotted the histogram of the conditional probabilities $P_U(\s_2|\s_1)*2^{21}$, which is normalized and sum to $1$. This also means that the XEB of the this distribution is exactly $0$. From the figure we can see that the histogram of the $2^{21}$ conditional probabilities coincides perfectly to the Porter-Thomas distribution. 

\end{document}